\documentclass[preprint,5p,times,twocolumn]{elsarticle}

\usepackage{float}
\usepackage{tcolorbox}

\newfloat{example}{thp}{exa}
\floatname{example}{Example}
\floatstyle{boxed}
\newcounter{exctr}

\tcbset{colback=yellow!10, bottom=10pt}

\usepackage{listings}
\usepackage{fancyvrb}
\usepackage{caption}
\usepackage{subcaption}
\usepackage{multicol}
\usepackage{comment}

\newcommand{\mywidth}{0.5\textwidth}

\providecommand{\logL}{$\log\!L$}				
\providecommand{\logZ}{$\log\!Z$}				
\providecommand{\tenlogZ}{$\log_{10}Z$}		
\providecommand{\logW}{$\log\!W$}				
\providecommand{\logLlow}{$\log\!L_{low}$}	
\providecommand{\Fxtheta}{$F(x\!:\!\theta)$}
\providecommand{\NestedSampler}{{\bf NestedSampler}}
\providecommand{\PhantomSampler}{{\bf PhantomSampler}}
\providecommand{\Explorer}{{\bf Explorer}}
\providecommand{\Problem}{{\bf Problem}}
\providecommand{\Sample}{{\bf Sample}}
\providecommand{\List}{{\bf List}}
\providecommand{\Walker}{{\bf Walker}}
\providecommand{\Model}{{\bf Model}}

\providecommand{\Base}{{\bf Base}}
\providecommand{\SimpleModel}{{SimpleModel}}
\providecommand{\Gauss}{{\bf Gauss}}
\providecommand{\Polynomial}{{\bf Polynomial}}
\providecommand{\RadialVelocity}{{\bf RadialVelocity}}
\providecommand{\Power}{{\bf Power}}

\providecommand{\Sine}{{\bf Sine}}
\providecommand{\Fixed}{{\bf Fixed}}

\providecommand{\Dynamic}{{\bf Dynamic}}
\providecommand{\Modifiable}{{\bf Modifiable}}
\providecommand{\Splines}{{\bf Splines}}
\providecommand{\Classic}{{\bf Classic}}
\providecommand{\ErrorsInXandY}{{\bf ErrorsInXandY}}
\providecommand{\MultipleOutput}{{\bf MultipleOutput}}
\providecommand{\Evidence}{{\bf Evidence}}
\providecommand{\Order}{{\bf Order}}
\providecommand{\Distribution}{{\bf Distribution}}
\providecommand{\ErrorDistribution}{{\bf ErrorDistribution}}
\providecommand{\Poisson}{{\bf Poisson}}
\providecommand{\Laplace}{{\bf Laplace}}
\providecommand{\Uniform}{{\bf Uniform}}
\providecommand{\Bernoulli}{{\bf Bernoulli}}
\providecommand{\Exponential}{{\bf Exponential}}
\providecommand{\Mixed}{{\bf Mixed}}
\providecommand{\Prior}{{\bf Prior}}
\providecommand{\Jeffreys}{{\bf Jeffreys}}
\providecommand{\Cauchy}{{\bf Cauchy}}
\providecommand{\Engine}{{\bf Engine}}
\providecommand{\Gibbs}{{\bf Gibbs}}
\providecommand{\Step}{{\bf Step}}
\providecommand{\Galilean}{{\bf Galilean}}
\providecommand{\Chord}{{\bf Chord}}
\providecommand{\Birth}{{\bf Birth}}
\providecommand{\Death}{{\bf Death}}
\providecommand{\Structure}{{\bf Structure}}
\providecommand{\Fitter}{{\bf Fitter}}
\providecommand{\QR}{{\bf QR}}
\providecommand{\LevenbergMarquardt}{{\bf LevenbergMarquardt}}
\providecommand{\Annealing}{{\bf Annealing}}
\providecommand{\Amoeba}{{\bf Amoeba}}
\providecommand{\Curve}{{\bf Curve}}
\providecommand{\ConvergenceError}{{\bf ConvergenceError}}
\providecommand{\result}{{\tt result()}}
\providecommand{\str}{{\tt \_\_str\_\_()}}
\providecommand{\init}{{\tt \_\_init\_\_()}}
\providecommand{\domaintounit}{{\tt domain2Unit()}}
\providecommand{\unittodomain}{{\tt unit2Domain()}}
\providecommand{\grow}{{\tt grow()}}
\providecommand{\shrink}{{\tt shrink()}}

\usepackage{lineno,hyperref}
\modulolinenumbers[5]

\journal{Astronomy and Computing}

\usepackage[figuresright]{rotating}

\begin{document}
\begin{frontmatter}

\title{BayesicFitting, a PYTHON Toolbox for Bayesian Fitting and Evidence Calculation.\\ \large{Including a Nested Sampling implementation}}


\author[sron]{Do Kester\corref{mycorrespondingauthor}}
\cortext[mycorrespondingauthor]{Corresponding author}
\ead{dokester@home.nl}

\author[sron,leiden]{Michael Mueller}

\address[sron]{SRON Netherlands Institute for Space Research, Postbus 800, 9700AV Groningen, The Netherlands}
\address[leiden]{Leiden Observatory, Leiden University, P.O. Box 9513, 2300 RA Leiden, The Netherlands}

\date{}

\begin{abstract}

BayesicFitting is a comprehensive, general-purpose toolbox for simple and standardized model fitting. Its fitting options range from simple least-squares methods, via maximum likelihood to fully Bayesian inference, working on a multitude of available models. BayesicFitting is open source and has been in development and use since the 1990s. It has been applied to a variety of science applications, chiefly in astronomy. 

BayesicFitting consists of a collection of PYTHON classes that can be combined to solve quite complicated inference problems. Amongst the classes are models, fitters, priors, error distributions, engines, samples, and of course  NestedSampler, our general-purpose implementation of the nested sampling algorithm. 

Nested sampling is a novel way to perform Bayesian calculations. It can be applied to inference problems, that consist of a parameterized model to fit measured data to. NestedSampler calculates the Bayesian evidence as the numeric integral over the posterior probability of (hyper)parameters of the problem. The solution in terms of the parameters is obtained as a set of weighted samples drawn from the posterior.

In this paper, we emphasize nested sampling and all classes that are directly connected to it. Additionally, we present the fitters, which fit the data by the least-squares method or the maximum likelihood method. They can also calculate the Bayesian evidence as a Gaussian approximation.

We will discuss the architecture of the toolbox. Which classes are present, what is their function, how they are related and implementational details where it gets complicated.

\end{abstract}



\end{frontmatter}

\section{Introduction}

\subsection{BayesicFitting}

The BayesicFitting software package is a PYTHON toolbox for doing Bayesian inference calculations in a simple and standardized way. Given a dataset and a parameterized model, the toolbox can optimize the parameters, obtain standard deviations, confidence regions, and most importantly calculate the evidence for the model. This evidence can be compared with the evidence for other models to decide which model fits best. {\bf The option to decide which model is better than another, lifts Bayesian fitting far above the traditional methods.}


There are several other Bayesian inference packages available to the scientific community: Bayesian packages in R~\cite{RCoreTeam:2017}, OpenBUGS~\cite{lunn:2009}, Emcee~\cite{emcee}, Stan~\cite{StanTeam}. Or specialized packages for X-ray astronomy: XSPEC~\cite{Arnaud:1996} and BXA~\cite{Buchner:2021}. However, they all miss one or more of the key elements of BayesicFitting: evidence calculation, generic applicability, nested sampling, and ready-made, combinable models. 

The BayesicFitting toolbox contains over 100 PYTHON classes, for models, fitters, priors, error distributions, etc. The models can be combined in various ways with each other to form new compound models. Models are discussed in section~\ref{sec:models}, preceded by the somewhat more abstract concept of problems in section~\ref{sec:problems}. The fitters are explained in section~\ref{sec:fitters}.
The sections in between are part of NestedSampler, mostly. They are devoted to  error distributions (section~\ref{sec:errordistribution}), priors (section~\ref{sec:priors}), engines (section~\ref{sec:engines}), and samples (section~\ref{sec:walkers}). After these sections, we discuss quality assurance in section~\ref{sec:quality}. And finally, we present a summary in section~\ref{sec:summary}.

While many fitters, including simple least-squares, are supported by BayesicFitting, the centerpiece of BayesicFitting (and the main focus of this paper) is NestedSampler. It is an independent implementation of the nested sampling algorithm developed by Skilling \cite{skilling:2004, skilling:2006}. NestedSampler is a fully Bayesian algorithm to calculate the evidence and obtain samples from the posterior. Nested sampling is explained in sections~\ref{sec:nestedsampling} and \ref{sec:nestedsampler}. However, we start with Bayes' rule in section~\ref{sec:bayes}.

Interspersed in the text are 3 examples on how to use the toolbox. They do not fit in the running text so they are presented as stand-alone sections called Example: a tea experiment, an exoplanet, and Boyle's law. 

\begin{example*}[!ht]
\begin{tcolorbox}
\par\noindent\rule{\textwidth}{0.8pt}
\refstepcounter{exctr}\label{ex:tea-1}
{\bf Example \arabic{exctr}: Julius' Tea Experiment}
\par\noindent\rule[6pt]{\textwidth}{0.4pt}
\begin{multicols}{2}
\setlength{\parindent}{10pt}

In a chemistry class, the students had to measure the transparency of water mixed with an increasing amount of strong tea. Julius, son of the first author, took data at 6 equally spaced dilutions between 0 and 100\% tea and measured the light passing through with an app on his phone.

We define a model that, from a physics point of view, should fit these data: an exponential on a constant background. This is a non-linear model so we need a non-linear fitter: LevenbergMarquardt.

\begin{Verbatim}[frame=single, fontsize=\small] 
 tea   = [ 0, 0.2, 0.4, 0.6, 0.8, 1.0]
 light = [9.55, 6.34, 5.13, 4.11, 3.17, 2.80]

 model = ExpModel() + PolynomialModel( 0 )
 fitter = LevenbergMarquardtFitter( tea, model )
 params = fitter.fit( light, plot=True )
 print( params )
 [ 7.18  -2.46  2.27]
 print( fitter.stdevs )
 [ 0.44   0.39  0.44]
\end{Verbatim}


\begin{center}
\includegraphics[width=0.4\textwidth, trim=0 0 0 0.3in]
{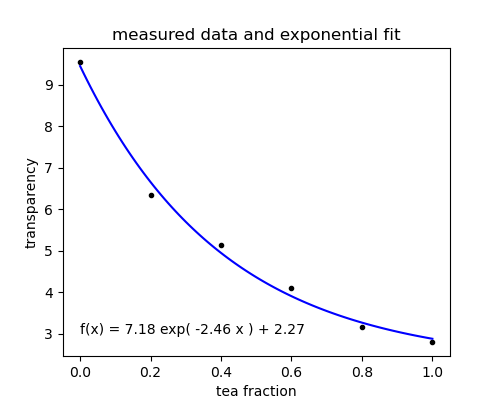}

\parbox{5cm}{\footnotesize Figure E1.1. Julius' tea experiment.}
\end{center}


The results of the fit are shown in figure E1.1. The equation that optimally fits the data is listed in the figure.

We can at least conclude that the experiment was executed properly, as the data and the expected model are in qualitative accordance.

\end{multicols}
\end{tcolorbox}
\end{example*}

\subsection{Pedigree}

BayesicFitting started life in the 1990s as a JAVA package inside the Herschel Common Software System (HCSS) \citep{ott:2010} of ESA's Herschel satellite. As a HIFI contribution to the system \cite{shipman:2017, edwards:2019}, we elected to write the fitter toolbox, building on 40 years of experience in data modeling for instrument calibration and other data analysis projects. 

At the end of Herschels Post-Operations period, maintenance of the HCSS system was suspended. 
Since this lack of maintenance will unavoidably lead to loss of functionality and eventually to loss of fitness, we decided to translate the toolbox into PYTHON, using NumPy \citep{harris2020array}, SciPy \citep{2020SciPy-NMeth}, and AstroPy \cite{price2018astropy} as basis for relevant and speedy array operations. A small program, j2py \citep{j2py:2014}, adapted and expanded for this purpose, was used in translating the JAVA code to PYTHON. It served mostly to keep in place the class structure with all its methods and inheritances and, very importantly, the documentation. Still, all code lines had to be inspected, corrected, and pythonized. By now BayesicFitting is larger and more powerful than the HCSS fitter toolbox ever was.

As part of HCSS, the fitter toolbox was used in the software pipelines of the 3 instruments onboard the Herschel satellite. The pipelines processed all data taken by Herschel whether astronomic observations or calibration measurements; even ground-based instruments tests needed to pass through the pipelines \cite{ott:2010} without failure. Consequently, the fitter toolbox was extensively tested in an astronomical environment. By carefully translating the JAVA-based fitter toolbox we introduced extra security into the new PYTHON-based BayesicFitting toolbox.

As an interpreted language PYTHON is not very fast. However, all array operations are relegated to the NumPy and SciPy packages which are written in C-code, compiled, and linked. Most of the CPU is spent while calculating the likelihood function over the data arrays, in compiled code.
More speed can be obtained by linking a threaded version of NumPy, SciPy, and the underlying BLAS/LaPack libraries. None of the classes is using global constants or even class attributes, so each object is an independent entity.



\subsection{Notation}

BayesicFitting is written in PYTHON as a collection of classes. These classes have names in CamelCase. In this paper, classes are written in boldface. So \Model\ is a class inside BayesicFitting built around a mathematical model.
There is no sharp dividing line, however. Notations might mix.
Classes with similar endings belong to the same family tree of \Model s, \Engine s, \Prior s, etc. These latter ones are the central members of the hierarchy. Interaction with other classes is addressed through them.  

Actual PYTHON code is written in {\tt teletype}, either inline or in small boxed code blocks. Due to space limitations and for readability the PYTHON code is ``stylized''. E.g. the reference to {\tt self} is often omitted where it is obvious.

\subsection{Availability} 

BayesicFitting is released in the public domain under the GPL3 license. It can be obtained from

\begin{table}[!ht]
\begin{tabular}{ll}
PyPi   & https://pypi.org/project/BayesicFitting \\
GitHub & https://github.com/dokester/BayesicFitting  \\
zenodo & https://zenodo.org \\
       & doi:10.5281/zenodo.2597200.\\
home   & https://bayesicfitting.nl \\
\end{tabular}
\end{table}

\section{Bayes' Theorem}
\label{sec:bayes}

This is not the place to expound Bayesian Statistics and Inference. We refer the interested reader to textbooks such as \citep{sivia:2006, jaynes:2003, bishop:2006}.
On a fundamental level, Bayes' theorem is the reason we can say something about parameters when we only know something about data. 

In this vein, we introduce Bayes' theorem to update knowledge, as-is: a relation of (densities of) probabilities, $p$, about statements, $a$ and $b$ in the presence of background information, $I$.

\begin{equation}
p(a\,|\,b,I) = \frac{p(a\,|\,I) \, p(b\,|\,a,I)}{p(b\,|\,I)}
\label{eq:bayes-std}
\end{equation}
\noindent
In terms of model inference, this transforms into a relation of probability densities between parameters, $\theta$, data, $d$, and a model, $m$.

\begin{equation}
p(\theta\,|\,d, m) = \frac{p(\theta\,|\,m) \, p(d\,|\,\theta,m)}{p(d\,|\,m)}
\label{eq:bayes-inf}
\end{equation}

\noindent
Each of these elements has its own symbol and a name. In words, we can write

\begin{equation}
\mbox{posterior} = \frac{\mbox{prior} \times \mbox{likelihood}}{\mbox{evidence}}.
\label{eq:bayes-names}
\end{equation}

\noindent
As the prior and the posterior are both distributions we give them the same symbol $P$, where the prior gets a subscript of 0 to indicate it is the probability we start with. By multiplying it by the likelihood, $L$, and dividing by the evidence, $Z$, we obtain the final probability, $P$, the posterior 

\begin{equation}
P = \frac{P_0 \times L}{Z}
\label{eq:bayes-symbol}
\end{equation}

\noindent
To emphasize that Bayes' rule updates our knowledge, we can borrow notation from computer languages.

\begin{equation}
P\,*\!\!= L\,/\,Z
\label{eq:bayes-update}
\end{equation}

\noindent
It also underlines that the posterior of one problem can be the prior of the next.\footnote{This is one of those things that are mathematically correct, but computationally very hard to accomplish, except for toy problems.}

\begin{description}
\item[posterior] The posterior distribution is a multi-dimensional distribution of the parameters given the model and the data. The posterior represents our updated knowledge of the parameters after taking into account the data. Parameter averages, standard deviations, and other items can be extracted from the posterior.
\item[prior] The prior distribution is the probability of the parameters given the model without the (present) data. They are educated guesses of the ranges over which the parameters of the model might vary. There is a lot of (intentional) confusion about priors. But only theoreticians can imagine a model where they (pretend to) have no knowledge at all about the extent of the parameters. Real practitioners never have that problem. There are always limits. And in most cases, the data overwhelm the specific choice of the priors anyway. If not, the data bear little information about the problem at hand, either by paucity or by irrelevance.
The prior could also be the result of a previous inference problem. Bayes' theorem updates knowledge.  
\item[likelihood] The likelihood is the calculated probability of the errors, the differences between data and model, given a set of parameters. The choice for the error distribution to be used depends on the problem itself and on the data. 
\item[evidence] On the face of it, the evidence is just a normalization factor, needed to have to probabilities of the posterior integrate to 1.0. And this is indeed how it is calculated:
\begin{equation}
Z = \int P_0\, L\, d\theta
\label{eq:bayes-Z}
\end{equation}
However, it can play a much more important role as the probability of the model given the data. Hence its name: evidence. It shows how good one model is with respect to another.
\end{description}
We write equation~\ref{eq:bayes-std} for model and data only; the probability of the model given the data.

\begin{equation}
p(m\,|\,d) = \frac{p(m) \, p(d\,|\,m)}{p(d)}
\label{eq:bayes-evi}
\end{equation}

The normalization factor, $p(d)$, cannot be assessed, as it would have us integrate over all possible models. However, if we restrict our universe to a small number of models, say 2, we can write the odds ratio between model 1 and model 2 as

\begin{equation}
\frac{p(m_1\,|\,d)}{p(m_2\,|\,d)} = \frac{p(m_1)}{p(m_2)} \frac{p(d\,|\,m_1)}{p(d\,|\,m_2)} 
\label{eq:bayes-rev}
\end{equation}

Assuming that we have no preference for either $m_1$ or $m_2$ i.e. the prior odds ratio is 1.0, we can determine which model is better than the other and by how much. The second factor, the evidence ratio, is also called the Bayes factor, $B_{12}$.

This possibility for favoring one model over another is the most important feature that lifts Bayesian inference over the more traditional versions of inference. The papers \citep{kester:1999, kester:2000, kester:2003, kester:2010a, kester:2010b, kester:2014} are all about problems that could only be solved by selecting the best model. And never, in our experience, did it fail Laplace's adage: ``Probability theory is nothing more than common sense reduced to calculation.''

\section{Nested Sampling}
\label{sec:nestedsampling}


Nested sampling is an algorithm developed by Skilling \cite{skilling:2004} designed to calculate the evidence, $Z$ of equation~\ref{eq:bayes-Z}. Unlike other Markov Chain Monte Carlo (MCMC) algorithms it does not need a ``burn-in'' and it has a well-defined stopping criterion. While calculating the evidence, nested sampling produces samples, randomly drawn from the posterior distribution.


The algorithm initializes an ensemble of N walkers, points in parameter space, randomly distributed over the prior. For all walkers, the log-likelihood, \logL, is calculated. 

In subsequent iterations the walker with the lowest likelihood, \logLlow\ is stored in a sample list, supplemented with a weight proportional to its contribution to the posterior. The walker is removed from the ensemble. A new walker is generated, randomly distributed over the prior, that has a \logL\ that is {\em higher} than the \logLlow\ of the discarded walker. This way we are climbing the likelihood mountain. 
As each walker represents 1/N part of the available remaining space we can numerically integrate the likelihood in successive steps. The prior is taken into account by the generation of the walkers via importance sampling. When the top of the likelihood is reached, the integral is complete: the evidence is obtained.  


The nested sampling algorithm can be explained in one small paragraph. Hereafter we will address our general-purpose implementation of nested sampling.

\section{NestedSampler}
\label{sec:nestedsampler}

Suppose we have a problem containing parameters and data that bear relevance to these parameters. The problem could be a classic regression problem or something more complicated, as addressed in section \ref{sec:problems}.

\NestedSampler\ takes the problem and applies the nested sampling algorithm to obtain the evidence, and posterior in the form of a list of samples. The algorithm, which is called with \NestedSampler s {\tt sample()} method, can be split into 7 steps. See Sivia~\citep{sivia:2006} for more details.
\begin{enumerate}
\item Initialize an ensemble of N walkers, randomly distributed over the  parameter space according to the priors. Calculate the \logL, the log of the likelihood for each of the walkers.

During the iterative process, the space occupied by the walkers shrinks on average as an exponential function
\begin{equation}
X(i) = \exp( -i / N )
\end{equation}
where $i$ is the iteration number. 
\item Find the walker with the lowest log-likelihood, \logLlow, the worst in the ensemble. 
The width for which the \logLlow\ is valid is the amount of shrinkage in the iteration. It equals the difference between subsequent $X$'s.
\begin{Verbatim}[frame=single, fontsize=\small]
 logWidth = log( X(i-1) - X(i) )
\end{Verbatim}
\item Copy the worst walker, properly weighted, to the list of samples. The weight of the sample is calculated as the product of the likelihood and the width of the area for which it is valid. 
\begin{Verbatim}[frame=single, fontsize=\small]
 logWgt = logL + logWidth
\end{Verbatim}
\item Update the developing integral for the evidence, Z, by adding the weight. As we do our calculations in the log we need the function {\tt numpy.logaddexp()}. Similarly, update the Kulback-Leibler information, H, also known as the Kulback-Leibler divergence. 

\begin{equation}
H \approx   \sum_i \frac{W_i}{Z} \log \frac{L_i}{Z}
\end{equation}

\begin{Verbatim}[frame=single, fontsize=\small]
 logZn = numpy.logaddexp( logZ, logWgt )
 # update Information, H
 H = ( math.exp( logWgt - logZn ) * logL +
       math.exp( logZ - logZn ) * 
       ( H + logZ ) - logZn )
 logZ = logZn
\end{Verbatim}
\item Remove the worst walker from the ensemble.
\item Duplicate a randomly selected walker from the remaining ensemble. The duplicated walker needs to be moved around while its \logL\ stays higher than that of the discarded walker until it is randomly distributed over the prior. As the other walkers were already randomly distributed and the new one is just made so, we have a new ensemble that is randomly distributed over a shrunken volume while all its \logL s are above the limit set in step 2. The exploration of the available space is done by the engines of section~\ref{sec:engines}. This step is the hardest of them all. It is called the ``Central Problem'' of nested sampling \cite{stokes:2017}.
\item Go back to step 2 until done. While the remaining space is slowly but exponentially decreasing, the likelihood is increasing until its maximum is reached. While the likelihood stays at its maximum, the available volume continues to go down. The product of these two counterbalancing trends produces a hump-like behavior in the contributions to the integral of the posterior. When the hump is passed and the contributions are back to (almost) zero it is time to stop. More strictly the iterations are stopped when
\begin{Verbatim}[frame=single, fontsize=\small]
 if iteration > 2 * N * H : 
     break
\end{Verbatim}
\end{enumerate}

When the iteration is stopped, the remaining walkers are copied to the list of samples. The {\tt sample()} method returns the evidence, or more specificly, the $\log_{10}$ of the evidence. This \tenlogZ, multiplied by 10, can be viewed as decibel when comparing models. 
From the samplelist, obtained as the attribute {\tt samples} from \NestedSampler, all kinds of statistics can be derived about the problem.

\begin{example*}[!ht]
\begin{tcolorbox}
\par\noindent\rule{\textwidth}{0.8pt}
\refstepcounter{exctr}\label{ex:exo}
{\bf Example \arabic{exctr}: Exoplanets around HD 2039}
\par\noindent\rule[6pt]{\textwidth}{0.4pt}
\begin{multicols}{2}
\setlength{\parindent}{10pt}

Tinney et al.~\citep{tinney:2003} have published radial velocity measurements of the star HD 2039, from which they conclude it has an exoplanet. Gregory~\citep{gregory:2005} reanalyzed the same data with a tentative assertion that there might be a second exoplanet. 

Lets use BayesicFitting to evaluate the evidence for one or two planets, given the data. However firstly, we investigate the case with no planets at all, assuming that the data can be explained by just a constant systemic velocity. Its evidence will be used as reference to the other cases.

\begin{Verbatim}[frame=single, fontsize=\small]
model = PolynomialModel( 0 )
model.setLimits( -100, 100 )
\end{Verbatim}

We start \NestedSampler\ with the model and the data: julian days ({\tt jd}) and radial velocity ({\tt rv}). Uncertainties were also published. We introduce them as weights, inverse square roots of the uncertainties. By default \NestedSampler\ calculates likelihood for a \Gauss\ErrorDistribution\ with an unknown scale factor that acts as a hyperparameter to the problem. It also needs a prior, a \Jeffreys\Prior\ with limits between (0.01, 100).

\begin{Verbatim}[frame=single, fontsize=\small]
ns = NestedSampler( jd, model, rv, weights=wgt )
ns.distribution.setLimits( [0.01,100] )
logZ = ns.sample()
\end{Verbatim}

The {\tt ns.sample()} method returns the $\log_{10}$ of the evidence: $\log Z_0 = -66.22$. The value will act as a reference in our next calculations.

The radial velocity algorithm we took from Gregory \citep{gregory:2005} and encoded it in \RadialVelocity\Model.  The model has 5 parameters: eccentricity, amplitude, period, longitude of the periastron, and time of the periastron. The first 3 parameters are positive by definition. Eccentricity needs to be $< 1$ for stable orbits; amplitude and period find their natural maximum in the data, max of {\tt rv} and {\tt jd} resp. They are assigned a \Uniform\Prior\ with limits as indicated. The longitude and time are phases: they need a circular \Uniform\Prior.

\begin{Verbatim}[frame=single, fontsize=\small]
rvm = RadialVelocityModel()
lolim = [0, 0, 0]
hilim = [1, 200, 1500]
rvm.setLimits( lolim, hilim )
pr = UniformPrior(circular=True,limits=[0,twopi])
rvm.setPrior( 3, pr )
rvm.setPrior( 4, pr )
model += rvm
\end{Verbatim}

Running \NestedSampler\ with this model yields the result displayed in figure~E2.1. 
The average parameters obtained from the posterior samples, are all equal within the error bars to the ones published in the earlier studies. 
With $\log Z_0$, we can calculated the odds the model with one exoplanet, $m_1$, to the model without planets, $m_0$, from equation~\ref{eq:bayes-rev}.
\begin{equation}
\frac{p( m_1\,|\,d )}{p(m_0\,|\,d)} = 5.9 \times 10^9
\end{equation}
So there is overwhelming evidence for a planet in the system.

{\centering
   \includegraphics[width=\mywidth, trim=0 0.3in 0 0]{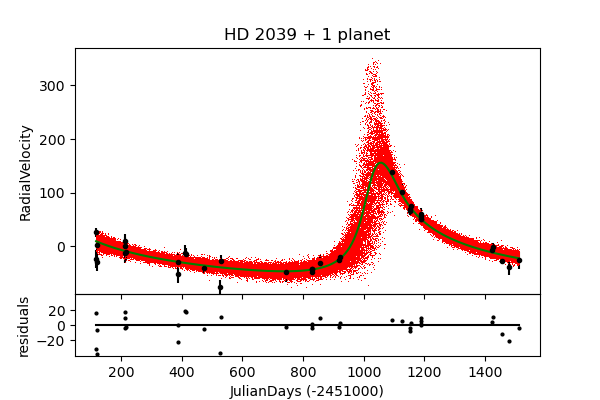}
   \label{fig:exo}
}
\begin{center}
\parbox{7cm}{\footnotesize Figure E2.1. The black point represent the data. The red dots are samples from the posterior for a model with one planet; the green line is the average of the posterior. In the lower panel the residuals are shown with respect to the average.}  
\end{center}

\vspace{5pt}

For the case of two planets we add another \RadialVelocity\Model\ to {\tt model}. All priors are the same except for the period. To avoid degeneracy of the orbits, we limit the prior for the period of planet B to uniform between [0, 1000]. 

\begin{Verbatim}[frame=single, fontsize=\small]
rvb = rvm.copy()
rvb.setPrior( 2, UniformPrior( limits=[0,1000] ))
model += rvb
\end{Verbatim}

When running the 2 planet model through \NestedSampler\ we obtain a Bayes factor $B_{12} = 0.1$, indicating that the 2 planets model is about 10 times less probable than the 1 planet model. 

Another indication that a second planet cannot be found in this data set, is the size of the standard deviations of its parameters; they let the parameters vary almost over the complete range of their priors. The data did not attribute to a sharper definition of the parameters. We did not learn much. Before application of the data we knew where the parameters were (i.e. within the priors) and after application they are still there.
\vspace{10pt}

\end{multicols}

\end{tcolorbox}
\end{example*}

\subsection{Explorer}

The \Explorer\ addresses step 6 in the list above. It calls the list of \Engine s in random order until enough randomness is achieved. This is a balance between efficiency and thoroughness. Efficiency for speedy calculations and thoroughness for losing correlation with the starting point. Experience showed that between 5 and 10 valid steps are enough.

The \Engine s can be put in separate threads if we are exploring more than one \Walker\ per iteration to parallelize and speed up the calculations.

\subsection{PhantomSampler}
\label{sec:phantom}

The \PhantomSampler\ is quite similar to the  \NestedSampler\ except that it uses part of the phantoms (see section~\ref{sec:phantoms}) to speed up the calculations. Each iteration several \Walker s with the lowest \logL, are removed from the ensemble and copied to the \Sample\List. Only one  \Walker\ is evolved, subject to a \logLlow\ equal to the best of the removed ones.
The other \Walker s are replaced with phantoms randomly chosen from the evolved path.  

\section{Problem}
\label{sec:problems}

A \Problem\ is a class that is defined by data, $D$, parameters, $\theta$, and a forward transform from parameters to mock data i.e. the data values according to the problem. From this, it should be possible to calculate a measure of distance between real data and mock data and from that the likelihood.

Data, parameters, and the transform are stored in the \Problem.
Optionally, weights can be stored too.
Throughout the BayesicFitting toolbox weights are defined as equivalent to a ``quantity of data''. A weight of $w$ is equivalent to having $w$ of the same data. Of course, a weight could be the inverse square of a standard deviation obtained earlier for the data-point, but it does not necessarily need to be. Weights are more general. E.g. a weight of 0 results in the omission of the data-point from the solution which is impossible when using standard deviations.

Traditionally the data consists of independent variables, $x$, which are known without error, and dependent variables, $d$. The transform is defined in a \Model\ class as a parameterized mathematical function, $F$. Given a set of values for the parameters, the function calculates mock data, $y =$~\Fxtheta. The mock data is the \Model s guess for $d$. The real data and the mock data go into the \ErrorDistribution\ to calculate the likelihood. 
The traditional case, i.e. only errors in the dependent variable, $y$, is called a \Classic\Problem. It is the default and it does not have to be declared. 

Another option is an \ErrorsInXandY\Problem\ where the $x$ variables are known with similar (un)certainty as the $d$ variables. An example in case is a color-color diagram in astronomy where there are color values on both axes; one not more certain than the other. We have to define extra unknown quantities, $u$, one for each value of $x$. The $u$ need to be estimated along with the parameters $\theta$. As we are not directly interested in them, they are called nuisance parameters. The nuisance parameters need priors too. Here we would advise using a \Gauss\Prior\ centered on the $x$ values. We don't want the $u$ to stray too much from the corresponding $x$. Both pairs, $(d,y)$, and $(u,x)$ go into the \ErrorDistribution\ to calculate the likelihood. In these problems, we have more parameters to estimate than there are data. In Bayesian inference, this is not a problem because the priors stabilize the solution.

In a \MultipleOutput\Problem, the data variable $d$ has more than 1 dimension. Such is e.g. the case in fitting stellar orbits as they appear on the sky. Or when $d$ is the outcome of a football match \citep{kester:2010a}. Obviously, the \Model\ needs to produce mock data of the same dimensionality as the data, $d$. The \MultipleOutput\Problem\ flattens the multi-dimensional output to a one-dimensional array, so it can be passed to the \ErrorDistribution. 

The final \Problem s we would like to discuss are discrete. The probabilities in equation~\ref{eq:bayes-inf} are not densities because the possible configurations are discrete, each having its own likelihood value. Consequently the integral of equation~\ref{eq:bayes-Z} reverts to a sum over all configurations. However, the number of options is so large that they cannot sensibly be enumerated. We treat the sum as an integral that ironically, is calculated by the \NestedSampler\ numerically, as a sum. 

The first discrete problem is the \Evidence\Problem. We lift the nested sampling to the next level. Instead of calculating the \logZ\ of equation~\ref{eq:bayes-inf}, we calculate equation~\ref{eq:bayes-evi}. As equation~\ref{eq:bayes-evi} only functions as a comparison between different models, we need a way to generate models. The \Model\ class has children which are \Dynamic\ and/or \Modifiable, see section~\ref{sec:dynamod}. They have a dynamic number of components and hence of parameters, or their internal structure can be modified at a fixed number of parameters. In an \Evidence\Problem\ the optimal number of components and internal structure is found along with its preference over other configurations.

\Order\Problem s try to find a certain order in the elements of its problem domain. E.g. a school schedule\footnote{The school scheduling software is not part of the BayesicFitting toolbox}; which lessons have to be presented by what teachers to what sets of pupils in which classrooms and when \citep{bontekoe:2006}. Probabilistic results do not work in this case. A singular ``best'' result is what we are looking for. 
Our choice for the likelihood is the exp of the negated cost function which weighs the misfits in the schedule. This likelihood is unnormalized as normalization would entail summation over all configurations, which we cannot do because of its multitude. The normalization factor will be taken up in the evidence. 

The traveling salesman also fits in this problem category, the cost function is now the length of the path traveled.

All these problems can be addressed in \NestedSampler, although to solve the \Order\Problem\ specialized \ErrorDistribution\ and \Engine s need to be written.

\section{Models}
\label{sec:models}

\begin{figure}[!ht]
   \includegraphics[width=\mywidth]{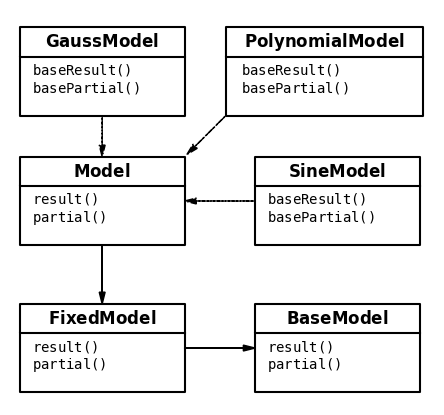}
   \caption{Schematic hierarchy of the model classes. Only a small selection of classes that inherit from \Model\ is shown. The arrow points to the parent class it is inheriting from.}
   \label{fig:models}
\end{figure}

The \Model s are a large hierarchy of classes centered around \Model\ itself. In figure~\ref{fig:models} the inheritance tree of models is sche\-matically shown. At the top of the tree \Gauss\Model, \Polynomial\Model\ and \Sine\Model\ are shown as representatives of all simple models that inherit from \Model. Collectively we will indicate them as \SimpleModel s. 

These \SimpleModel s inherit from \Model\ and subsequently from \Fixed\Model\ and \Base\Model. Actually, a \SimpleModel\ is mostly a specialization of \Base\Model\ at the root of the tree. However, as all interactions with other classes are organized via \Model, the inheritance line from \Base\Model\ to each \SimpleModel\ has to go through \Fixed\Model, \Model, and some auxiliary classes, not mentioned here. This way the \SimpleModel s are \Model s on their own and more complicated constructs built from \SimpleModel s are \Model s too.

\subsection{BaseModel}

The \Base\Model\ is an abstract (or base) class that should not be instantiated directly. It comes to life when a \SimpleModel\ is invoked. \Base\Model\ defines attributes and methods relevant to simple models. Attributes like the number of parameters, the dimensions of input and output, parameter names, whether some parameters need to be positive or non-zero. The priors on the parameters are stored here too as a list of \Prior s. And methods to act upon these attributes. 

Other methods like {\tt result()} and the partial derivative to the parameters {\tt partial()} directly call {\tt baseResult()} and {\tt basePartial()} in the relevant \SimpleModel. As is the \str\ method which directly calls the {\tt baseName()} method of its ultimate great-grandchild.

\subsection{SimpleModels}

A \SimpleModel\ is the concrete realization of the \Base\Model. It is built around a mathematical function, \Fxtheta, implementing its result, its partial derivatives to each of the parameters, and its derivative to $x$. There are more than 50 different \SimpleModel s in the toolkit. They can be used as-is or as building blocks for more complicated models as described below.

What exactly constitutes a simple function is a matter of taste and efficiency. Sometimes a more complicated function, which could in principle be constructed from building blocks, is better constructed as one class. At least it can be done more efficiently.

\subsection{FixedModel}

The \Fixed\Model\ is an abstract class too, in which one or more parameters is replaced by either a fixed, constant number or by the results of another \Model. The fixed model loses one parameter for each fixed one and possibly gains the parameters of the replacing models, which are placed at the end of the parameter list. 

When replacing a parameter with the results of another model there are no broadcasting problems as the length of the output arrays, $y$, are equal to those of the input, $x$. 
Partials and derivatives are calculated using the chain rule. 

A simple model can only be ``fixed'' at its construction and it remains fixed during its lifetime.
For all practical purposes a \Fixed\Model\ acts as a single unit: a simple model.

\begin{Verbatim}[frame=single, fontsize=\small]
 print( PolynomialModel( 2 ) )
 Polynomial: f(x:p) = p_0 + p_1 * x + p_2 * x^2
 print( PolynomialModel( 2, fixed={1:1.5} ) )
 Polynomial: f(x:p) = p_0 + 1.5 * x + p_1 * x^2
\end{Verbatim}

The \Polynomial\Model\ of order 2, has 3 parameters. When we fix the slope at 1.5, it has 2 parameters left: for the constant and for the $x^2$. 

\subsection{Compound Models}

Every \Model\ has a link either to {\tt None} or to another \Model. {\tt None} indicates the end of the linked list. Upon creation, all \SimpleModel s have a {\tt None} link.  

Linked \Model s have an associated list of operators, one for each link, which tells how the results of the next model operate upon the previously obtained results. Operators can be one of $+$ (addition), $-$ (subtraction), $*$ (multiplication), $/$ (division) or $|$ (pipe). The first 4 do the obvious thing: they add, subtract multiply or divide the results of the two \Model s to obtain a new result. The pipe uses the results obtained by the \Model\ on the left-hand side, as input of the \Model\ on the right-hand side. Like a UNIX pipe.

\begin{figure}[!ht]
   \includegraphics[width=\mywidth]{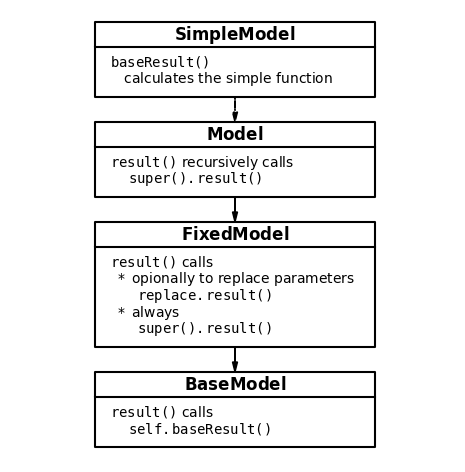}
   \caption{Calling sequence of the \result\ method, explained in the text below.}
   \label{fig:results}
\end{figure}

The linked list is stepped through in a recursive way for all methods that need information from each of the constituent \Model s. As an example, we show in figure~\ref{fig:results} how results for a compound model are obtained.

When the method \result\ is called in a compound model, it starts at the first \SimpleModel\ in the chain. A \SimpleModel\ does not have a \result\ method, so it falls through to \Model. The \result\ method in \Model\ recursively addresses the elements of its linked list. For each element, starting at itself, it calls the \result\ method in \Fixed\Model. \Fixed\Model\ starts looking for parameters that might be replaced by another \Model. This might be a full-fledged compound model so it calls \result\ in another \Model. When the results have returned it looks for other parameters that might have been replaced by constants and calls \result\ in \Base\Model\ with the reconstructed parameters. This last method directly calls {\tt baseResult()} in the relevant \SimpleModel. All recursively found results are operated upon according to their associated operators.

Two linked \Model s form a new unit which has (virtual) brackets around the operation. Its compound results are obtained before progressing down the chain. 
This might result in counterintuitive results that ignore normal operator preferences. E.g. {\tt m1 + m2 * m3} is evaluated as {\tt (m1 + m2) * m3}. 
However by reordering the chain or inserting extra brackets or separate evaluation of a subchain the intuitive outcome can be obtained. 

\begin{Verbatim}[frame=single, fontsize=\small]
 m4 = m2 * m3             # m4 is a new unit
 mdl1 = m1 + m4           # == ( m1 + ( m2 * m3 ))
 mdl2 = m2 * m3 + m1      # equivalent to mdl1
 mdl3 = m1 + ( m2 * m3 )  # equivalent to mdl1
\end{Verbatim}


\subsection{Dynamic and Modifiable}
\label{sec:dynamod}

\Dynamic\ and \Modifiable\ are base classes that can be used as extra parent classes for \Model s that can change its number of parameters or modify its internal structure. As an example we have a splines model where the number of knots can be changed (\Dynamic) and/or the location of the knots can be modified (\Modifiable). 

Upon instantiation of any \SimpleModel\ the init method of its parent is called via {\tt super().}\init. For a dynamic model that inherits from 2 (or 3) parent classes, it is not obvious which class {\tt super()} points to. The parent classes need to be called by name.

\begin{Verbatim}[frame=single, fontsize=\small]
 class SomeModel( Modifiable, Dynamic, Model ):
     def __init__( self ) :
         Modifiable.__init__()
         Dynamic.__init__()
         Model.__init__()
\end{Verbatim}

\section{ErrorDistribution}
\label{sec:errordistribution}

When we have selected a model and a set of (trial) values for the parameters we can calculate mock data, $y =$ \Fxtheta. We manipulate the parameter values such that mock data and measured data are as close as possible. Historically how to measure this closeness, was not all that obvious. Gauss minimized the squared distances, giving rise to the well known ``least-squares'' (or $\chi^2$) method. Laplace favored minimizing the absolute value of the distances. Both are valid for certain kinds of errors. As the least-squares methods proved easier to calculate it was for a long time the method of choice. 

In probability theory, we want to obtain the probabilities for the distances between mock and measured data. 
Error distributions will calculate the probability of these distances, resulting in the likelihood term in equation~\ref{eq:bayes-inf}: $p( d\,|\,\theta, m )$.

When we use the normal (or Gaussian) probability density function as the distribution of the errors, we get for each individual data point, $x_k$, and optionally an associated weight, $w_k$, the
calculated probability of its residual.

\begin{equation}
\mathcal{L}( x_k ) = \sqrt {\frac{w_k}{2\pi\sigma^2}} \exp\left( - 0.5 w_k \left(\frac{y_k - d_k}{\sigma}\right)^2 \right)
\label{eq:gauss}
\end{equation}

The $\sigma$ is a parameter of the error distribution and can be estimated independently as a hyperparameter of the problem. The optimal $\sigma$ represents the noise scale, i.e. the standard deviation of the residuals: data minus optimal model.

If the errors are independent of each other, the likelihood is the product of the individual contributions over all data points.

\begin{equation}
L = \prod_k \mathcal{L}( x_k )
\label{eq:prodL}
\end{equation}

As we perform our calculations in the log, equation~\ref{eq:prodL} reverts to the sum over the log of the individual contributions.
The log of the (normal) likelihood of the errors is equal to the sum of the squares of the errors. Maximizing the likelihood is equivalent to minimizing the sum of the squares, or equivalent to the least-squares method.

So we see that least-squares is an approximation of Bayesian fitting. 
When we would have taken the Laplace probability distribution as likelihood, we would have obtained something equivalent to Laplace' method of minimizing the absolute deviations. His method is a Bayesian approximation too.

Conversely, with the BayesicFitting toolbox, these traditional fittings can be done as well.

\subsection{ErrorDistribution classes}

Central in the tree of error distribution classes is \ErrorDistribution. It stores hyperparameters that might be present and might be added to the list of parameters to be estimated. These hyperparameters of course have their own \Prior s. \ErrorDistribution s act upon a \Problem\ and parameters, $\Theta$ which  includes the parameters of the model, $\theta$, and the hyperparameters of the error distribution.

Which particular error distribution is needed, is a choice of the practitioner depending on the experiment at hand. However, this is not the place to discuss these choices. We only present the options. 

For counting problems, we provide the \Poisson\ErrorDistribution. For median-like solutions with the 1-norm, there is the \Laplace\ErrorDistribution, for the 2-norm, the good old \Gauss\ErrorDistribution, equivalent to least-squares and for the $\infty$-norm we supply the \Uniform\ErrorDistribution\ where the maximum of the residuals is minimized. 
Other norms can be made with the \Exponential\ErrorDistribution, setting the power as norm. This power is an extra hyperparameter in addition to the scale hyperparameter that all norm-ed error distributions have. 
For categorical problems, we have the \Bernoulli\ErrorDistribution. 

And finally, for \Evidence\Problem s, there is the \Model\-\Distribution. 
It provides the likelihood term in equation~\ref{eq:bayes-evi}, $p(d\,|\,m)$, which is the same as the evidence term in equation~\ref{eq:bayes-inf}. \Model\Distribution\ calculates the evidence either by running \NestedSampler\ or as a Gaussian approximation using one of the \Fitter s.

\subsection{Mixed Error Distributions}

A \Mixed\ErrorDistribution\ combines the weighted contributions of two other distributions, either of the same class or different. It can be used in situations where there are spurious outliers, excursions much larger than the others. In such cases, equation~\ref{eq:prodL} is changed into

\begin{equation}
L = \prod \, \lambda \mathcal{L}_1( x_k ) + (1-\lambda) \mathcal{L}_2 ( x_k ) 
\label{eq:mixed}
\end{equation}

Again, as the individual contributions are only known in the log we have to use the {\tt logaddexp()} method from NumPy. The mixing factor, $\lambda$, can have values between 0 and 1. It is another hyperparameter of the \Mixed\ErrorDistribution\ next to the ones from both constituent distributions.

Because of these requirements in \Mixed\ErrorDistribution, the summation over the log-likelihood contributions of each data point is delegated to the base class \ErrorDistribution. 

\subsection{Constraints}
\label{sec:constrain}

The space available to the parameters is restricted by their priors. By construction the parameters are orthogonal. However, in real life, restrictions on combinations of parameters are also possible. In these cases, a constrain method needs to be defined which excludes certain regions of the priors (exclusion sampling).

\subsection{Cost Functions}

Nested sampling is a very general method to integrate a multidimensional exponentiated function. It does not necessarily need to be a likelihood function. It could be the length of the travels of the salesman or the number of misfits in a school schedule \citep{bontekoe:2006}. The likelihood proxy can be each of these quantities in the negative and exponentiated. Generally, these are known as cost functions. Cost functions are the unnormalized likelihoods of \Order\Problem s, see section~\ref{sec:problems}.

In \Order\Problem s the domain of the parameters is not continuous; it consists of a (large) number of options. N! in case of an N city route and for a typical Dutch secondary school it is $10^{3800}$. Consequently, there is no guarantee that nearby the copied point a valid new one can be found. We have to search for new \Walker s until we find a valid one.

\section{Priors}
\label{sec:priors}

To calculate the evidence we need to integrate the product of likelihood and prior over the parameters. Nested sampling performs the calculation by smart sampling of the posterior. The prior is taken into account by importance sampling. 

For this purpose all priors need to have a pair of methods \domaintounit, the cumulative distribution function (CDF) and \unittodomain, its inverse (iCDF). When one takes a random value, uniformly between 0 and 1, the inverse CDF transforms it into a random value from the prior. And the CDF can transform any parameter value back into unit space.

The continuity of CDF and iCDF ensure that we can sample from the prior in the vicinity of a given parameter value. We transform the boundaries of the vicinity to the unit range, find a uniformly distributed value within the unit boundaries and transform that value back to the parameter domain. We need this in the next section~\ref{sec:engines} about engines.

\subsection{Limits}

Some priors need limits to properly integrate to 1.0, like \Uniform\Prior\ and \Jeffreys\Prior. Others might be limited by choice. This limitation is handled in \Prior. When limits to the priors are provided we define in the unlimited domain the minimum and range in the unit-space, obtained from the {\tt lowLimit} and {\tt highLimit} in domain-space.

\begin{Verbatim}[frame=single, fontsize=\small]
 umin = domain2Unit( lowLimit )
 uran = domain2Unit( highLimit ) - umin
\end{Verbatim}

We manipulate the calling order of \unittodomain\ and \domaintounit\ methods which resolve to the methods in the specific prior, such that they first pass through the {\tt limited} versions.

\begin{Verbatim}[frame=single, fontsize=\small]
 self.baseDomain2Unit = self.domain2Unit
 self.baseUnit2Domain = self.unit2Domain
 self.domain2Unit = self.limitedDomain2Unit
 self.unit2Domain = self.limitedUnit2Domain
\end{Verbatim}

Now the methods \domaintounit\ or \unittodomain\ are renamed to their {\tt base} variant and calls to \domaintounit\ or \unittodomain\ will resolve to the {\tt limited} versions.

\begin{Verbatim}[frame=single, fontsize=\small]
 def limitedDomain2Unit( self, dval ) :
    return ( baseDomain2Unit( dval ) - umin ) / uran

 def limitedUnit2Domain( self, uval ) :
    return baseUnit2Domain( uval * uran + umin )
\end{Verbatim}

After this manipulation, any call to \domaintounit\ goes to {\tt limitedDomain2Unit()}\ in \Prior, which subsequently calls {\tt baseDomain2Unit()}, a renaming of \domaintounit\ in the specific prior. Similar for \unittodomain. 

\subsection{Circular}

Some parameters are by nature periodic, like phases. They need a circular prior. This is also arranged in \Prior. Any prior which is symmetric, $p(\mu-x) = p(\mu+x)$ for some central value, $\mu$, can be made circular. 

\begin{Verbatim}[frame=single, fontsize=\small]
 self.domain2Unit = self.circularDomain2Unit
 self.unit2Domain = self.circularUnit2Domain

 def circularDomain2Unit( self, dval ) :
    return ( limitedDomain2Unit( dval ) + 1 ) / 3

 def circularUnit2Domain( self, uval ) :
    return limitedUnit2Domain( ( uval * 3 ) % 1 )
\end{Verbatim}

The first method places the unit domain in a [1/3,2/3] box, so excursions over the edges are possible, while the second method expands it to [0,3] and then only the fractional part is kept. Overflow and underflow out of the original unit box is circled to the other side. 
The edges i.e. the limits, need to be set so that the prior value at the low limit is continuous to the prior value at the high limit.

\subsection{Computation}

Priors that have in theory an infinite domain are nonetheless confined in practice. These priors need to die out toward infinity, some faster than others. At a certain moment, the values become indistinguishable from 0. In table~\ref{tab:limits}, the computational boundaries of the domain are given for a number of \Prior s. When x (here the scaled distance to the center) is larger than the values given, the \Prior\ is essentially 0. No steps can be taken beyond them, so no solution will be found there.

\begin{table}[!ht]
\centering
\begin{tabular}{|l|c|c|}
\hline
\Prior  &  math  & boundaries\rule{0mm}{4mm} \\ 
\hline
\Gauss\Prior & $\exp( - 0.5 x^2 )$ &  [-6, 6]\rule{0mm}{4mm} \\
\Laplace\Prior & $\exp( - |x| )$ & [-36, 36] \\
\Exponential\Prior & $\exp( -x )$ & [0, 36] \\
\Cauchy\Prior & $1 /( 1 + x^2 )$ & [-1e16, +1e16] \\
\hline
\end{tabular}
\caption{Computational boundaries for some unbound \Prior s using 8 byte floats.}
\label{tab:limits}
\end{table}

Attractive as \Gauss\Prior s are as conjugate to the normal distribution, they are also dangerous. They can only be explored to 6 unit lengths from its center. This is not a mathematical limitation, but
a computational one caused by the use of 8 byte floats in normal present day computers

\section{Engines}
\label{sec:engines}

The \Engine s are the classes where the \Walker s are moved around according to their \Prior\ until they have found a new random position, that is still higher than \logLlow. It is easily said but much harder to accomplish. Good exploration is needed to get a solid estimate of the evidence integral. When exploring too much in the high likelihood regions the evidence will be too high. And when too much time is spent in the outskirts the evidence is too low. 

All \Engine s transform the parameters of a \Walker\ that need to be randomized, to unit space, using the \domaintounit\ method from the \Prior. They select steps in unit space in a uniformly random way. The \Prior\ method \unittodomain\ is used to transform the steps to the parameter domain. This way all steps are always distributed according to the \Prior\ of the parameter. 

We determine the step size by transforming
the maximum and minimum values of the parameters, taken over the ensemble to unit space. There it forms a rectangular container box that encompasses the present ensemble. The box is used as a proxy for the available space in which can be stepped. The size of the steps scales with this.

As we start our \Walker\ by copying a valid one and the likelihood is continuous, we can always find a new position nearby or far.

\begin{figure*}[!ht]
     \begin{minipage}{1.0\columnwidth}
         \centering
         \includegraphics[width=0.8\textwidth]{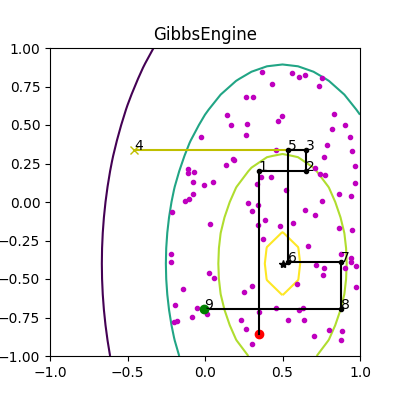}
     \end{minipage}
     \hfill{}
     \begin{minipage}{1.0\columnwidth}
         \centering
         \includegraphics[width=0.8\textwidth]{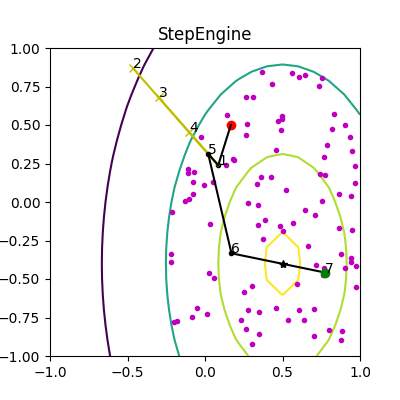}
     \end{minipage}
     \begin{minipage}{1.0\columnwidth}
         \centering
         \includegraphics[width=0.8\textwidth]{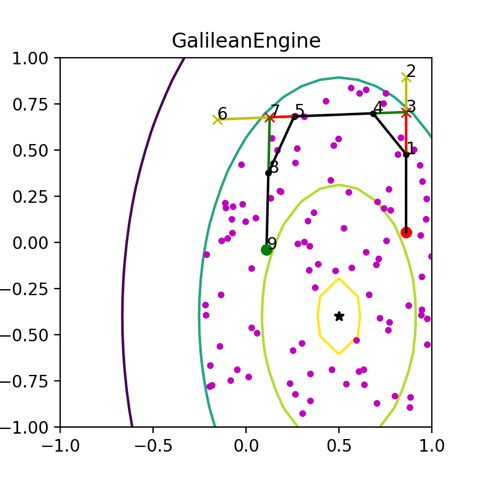}
     \end{minipage}
     \hfill{}
     \begin{minipage}{1.0\columnwidth}
         \centering
         \includegraphics[width=0.8\textwidth]{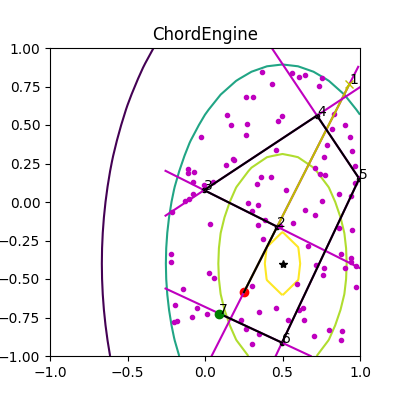}
     \end{minipage}
     \caption{Operation of 4 \Engine s on an arbitrary \Polynomial\Model\ of order 1. The axes are the 2 parameters of the problem, both with \Uniform\Prior s between -1 and 1. In all figures the contour of \logLlow is the mint one that encompasses the complete ensemble the 100 purple points. The engine starts at the red point and steps via the numbered points to the final green one. The black points are phantoms. The yellow excursions outside of the mint countour are failed points.}
\label{fig:engines}
\end{figure*}

In figure~\ref{fig:engines} one operation of the 4 engines discussed here, is displayed. It regards a simple linear \Problem\ with only 2 parameters for easy display. An ensemble of 100 \Walker s is shown as purple dots. For reference, some contours of equal \logL\ are shown. The one in mint encompassing the whole ensemble, is the present contour of \logLlow. Of course, the exact location of these contours is not known at any time during the operation of the \Engine. Here it is shown for reference only. Each \Engine\ starts at a copy of an existing \Walker, shown as the big red dot. The \Engine\ moves the point in (numbered) steps to its final position, the big green dot. Black dots mark acceptable (valid) steps as they are within the \logLlow\ contour. Yellow dots are steps outside the contour and are rejected (invalid). Details on the individual \Engine s are discussed below.

For all \Engine s it is of paramount importance never to accept an invalid location, not even temporarily. Once outside it can be very difficult the get back inside, especially in a high-dimensional parameter space.

\subsection{GibbsEngine}

Traditionally there is Gibbs sampling, here implemented as \Gibbs\Engine. It moves one parameter at a time by a random amount. If the new likelihood is above \logLlow\ the step is valid and the \Walker\ moves on. If not, the step size is decreased until it is accepted. The downside of Gibbs sampling is that is moves one parameter at a time and it performs a random walk. Performance scales with the number of parameters and it moves slowly from its original position. In 2 dimensions, as shown in figure~\ref{fig:engines}, it looks fine. The disadvantages don't show much. In more dimensions the parameters are addressed in random order; in two dimensions that does not work nicely.

\subsection{StepEngine}

A slightly more efficient way is the \Step\Engine. It also does a random walk but it moves in a random direction. All parameters are stepped at the same time. When the step goes outside the valid area, like step 2 in figure~\ref{fig:engines} it is decreased, first to 3 then to 4, and finally to 5 where it finds a valid position. In this example, the \Step\Engine\ looks efficient and simple. In much more dimensions the chance to step outside increases when we take steps of the size shown here. We have to revert to more, smaller steps. However, it is still a random walk that progresses by the square root of the number of steps.

\subsection{GalileanEngine}

The \Galilean\Engine\ is an independent implementation of an algorithm by the same name, designed by Skilling \cite{skilling:2012}. It
tries to amend the disadvantages of both the \Gibbs\Engine\ and the \Step\Engine. It selects a step in a random direction vector, $v$, and proceeds along that direction for as long as it is successful. When ultimately a step fails ($<$\logLlow) it tries to mirror on the \logL -surface to get back into valid space. In figure~\ref{fig:engines} position 2 (and also 6) finds itself outside the valid area defined by \logLlow. If it mirrors at position 2, it might not return into valid space. However, when we withdraw to position 1 to mirror, we are in danger of undersampling the border regions. 

To approximate the position of the \logLlow\ surface on the line (1,2), we interpolate from the \logL\ at 1 to the \logL\ at 2, at \logLlow to obtain position 3. This interpolation may not yield the exact location  \logLlow, it is closer. There is more chance that mirroring will indeed get us back into valid space. For mirroring we need to alter the present directional vector, $v$. 

\begin{equation}
v^{\prime} = v - 2 dL \frac{(dL.v)}{|dL|}
\end{equation}  

$v^{\prime}$ is the new stepping vector and $dL$ is the partial derivative of \logL\ to the parameters, $dL = \partial \log L / \partial \Theta$; and $(dL.v)$ is the inner product with $v$. The part of the step outside valid space, is taken in the mirrored direction, here to 4 (and 8). Stepping is continued in the new direction.

When mirroring is not successful we have to back up and retrace our steps in the reverse direction: $v^\prime = -v$. To all steps a 10\% random disturbance is added, if only to avoid exactly retracing our steps back to the starting point.

The \Galilean\Engine\ is the engine of choice for intermediate and large-sized problems. It has none of the disadvantages of \Gibbs\Engine\ or \Step\Engine. It can take small steps but as it moves in one direction it quickly traverses the available space to find a new random spot. The only disadvantage is that it needs the partial of \logL. All existing \Model s and \ErrorDistribution s have built-in methods for calculating analytical partials. When some future implementation omits these analytical partials, numeric calculations will automatically kick in.

\subsection{ChordEngine}

The \Chord\Engine\ is an independent implementation of the Polychord algorithm as described in \cite{handley:2015} or the Hit-and-Run algorithm of \cite{stokes:2017}. In unit space it draws a randomly oriented line through the present point and extends it until it reaches outside the valid space, yielding a start point and an endpoint. In figure~\ref{fig:engines} the purple lines extending from each point reach into invalid space or until the edge of the unit box. On this segment, a random point is selected. When it is valid ($>$\logLlow) it is the new point. Otherwise, either the start point or the endpoint is replaced by the new invalid point. And we select another random point on the new segment. Until we have a valid point. 

In figure~\ref{fig:engines} the first point on the random line through the starting point (red), turns out to be invalid after which we fall back to point 2 in valid space. Subsequent lines are made orthogonal to each other for as long as that is possible. In this 2-dimensional problem, the lines are orthogonal at even points only. At each next point, the possibilities for orthogonality are exhausted and a new random direction is found. 
  
The \Chord\Engine\ does not need the partial derivatives to \logL\ which is a bonus. However, it needs to extend the lines until they are in invalid space which in principle are extra likelihood calculations. However, our knowledge of the container box of the present ensemble helps here. We assume that once the line is outside the box, it reaches into invalid space, without further checking.

By default, \NestedSampler\ employs the \Galilean\Engine\ and the \Chord\Engine\ in a random order.

\subsection{BirthEngine and DeathEngine}

\Dynamic\ \Model s need extra \Engine s to increase or decrease the complexity of the model which manifests itself in the number of parameters the model has. In general, a change in the number of parameters is a discontinuous step. In some cases, for special values of the new parameter(s) the transition can have the same \logL\ as before and it can develop from there. 
An example here is a polynomial model. Increasing the order, but keeping the last parameter at 0, will yield the same \logL. Similarly with decreasing the order and having a last parameter of (almost) 0. So smooth transitions between these dynamic models exist. In other cases where this continuity does not exist, one can get stuck in a configuration that is not optimal.

All \Dynamic\ \Model s need methods \grow\ and \shrink\ that are called in the \Birth\Engine\ and \Death\Engine\ resp. The engines are added to the list of engines to be called by the \Explorer. The growing or shrinking of a model is governed by a \Prior, the growth prior, set in the \Model.

\subsection{StructureEngine}

\Modifiable\ \Model s have a method {\tt vary()} to alter the internal structure of the model, The method is called by the \Structure\Engine, which is automatically added to the list of engines when applicable. An example is a modifiable splines model where the location of one or more knots can be shifted. This change is a continuous one, contrary to the other option of dynamically altering the number of knots.

\subsection{Phantoms}
\label{sec:phantoms}

The intermediate, valid steps that a \Walker\ takes on its way to a new location are called Phantoms. (All black dots in figures~\ref{fig:engines}). For one \Walker\ its ``own'' phantoms are not very randomly distributed as they are on its path from its original position to the final one. However seen from a distance after many \Walker\ moves, they get disconnected and look more and more random. As there are many more phantoms than walkers, smart use of the phantoms might improve the speed of nested sampling. See section~\ref{sec:phantom}.   

\section{Walkers and Samples}
\label{sec:walkers}

The ensemble of \NestedSampler\ consists of \Walker s. Next to some administrative items a \Walker\ contains (a reference to) the problem, the parameters and hyperparameters, and its \logL.

At the end of its life when it is worst in the ensemble, the \Walker\ is converted into a \Sample\ and put into a \Sample\List. At that instance, a weight (stored as \logW) is added to the \Sample. This \Sample\List\ is a proxy for the posterior distribution of equation~\ref{eq:bayes-inf}. When nested sampling is finished, the weights are rescaled such that the sum of the weights is 1.0.  Weights are used to weigh the items extracted from the posterior (i.e. the \Sample\List).

\section{Fitters}
\label{sec:fitters}

Least-squares and maximum likelihood methods are encoded in \Fitter s of diverse pedigree. We already argued that these methods are approximations of Bayesian inference. We can take it even further. Provided that the priors of the parameters are uniform and the noise scale has a Jeffreys prior, both with limits, we can calculate the evidence as the integral over a Gaussian approximation of the posterior \cite{gull:1988}. 

As long as we compare approximations of similar models, like polynomials of different orders, sinusoidal components in a signal, etc. the evidences we obtain are very well comparable. We have used it massively and without failure in the past \cite{kester:1999}.  

We would not advise comparing a Gauss-approximated evidence with one obtained from nested sampling. There could easily creep in a bias stemming from non-gaussity of the posterior. 
A bias that would arguably fall in the same direction when similar approximations are taken. 

\subsection{BaseFitter}

The Gaussian approximation of the evidence is located in the root class of the fitter tree, \Base\Fitter. Other calculations that derive from the same approximation of the posterior, are obtained in this base class too. Amongst them the noise scale, standard deviations of the parameters, the covariance matrix, and confidence regions. 

\Base\Fitter\ checks the data (and weights) arrays for infinities and NaNs and issues an error when they are found.  

\subsection{Least Squares Fitters}

The most simple application of the least-squares method is for models, that are linear in its parameters. These linear problems can be solved by one pseudo inversion of the Hessian matrix. It is simple, fast, and reliable. The likelihood landscape for linear models is unimodal; there is only one minimum that can be found in one step.
Moreover the posterior for the parameters is Gaussian, provided that the prior is conjugate (\Gauss\Prior), or a \Uniform\Prior\ of sufficient width. In these cases the Gaussian ``approximation'' is exact. Linear fitters include \Fitter\ which encapsulates the {\tt numpy.linalg.solve()} method and \QR\Fitter\ that is built on the {\tt numpy.linalg.qr()} decomposition.

Non-linear problems can also be solved by the least-squares method, applied in an iterative way. There are several ways to crawl downhill in the $\chi^2$-landscape. The \LevenbergMarquardt\Fitter\ implements an algorithm using the gradient of $\chi^2$.  It proceeds downhill as long as it can, like a rolling stone, falling into the first minimum it encounters. There is no guarantee that this is the global minimum we are looking for. A smart choice of initial parameters is needed to end in the global minimum. 
\Curve\Fitter\ encapsulates the {\tt scipy.optimize.curve\_fit()} method; it contains another incarnation of Levenberg-Marquardt.

\subsection{Maximum Likelihood Fitters}

Maximum likelihood fitters are built around a ``minimizer'', either 
\Annealing\Amoeba\ or one of the 10 methods present in {\tt scipy.optimize.minimize()}. They all minimize the negative log of the likelihood. 
\Annealing\Amoeba\ employs a simplex that crawls downhill as in the Nelder-Mead method. However, to escape from local minima, it can apply simulated annealing at the cost of much longer runtimes. The fitter implementing this is \Amoeba\Fitter.
The scipy fitters employ various methods that are beyond the scope of this paper.

\section{Quality Assurance}
\label{sec:quality}

\subsection{Documentation}

The BayesicFitting package has a documentation section \cite{docs:2021} where several documents are available that can help the user to find its way within the BayesicFitting toolbox.
\begin{table}[!ht]
\begin{tabular}{ll}
Manual   & explaining the BayesicFitting toolbox in detail for\\
		 & easy usage \\
Design   & architectural design document. \\
Glossary & list of terms used in the toolbox and documents. \\
Style    & short note on coding style, that we adhere to.\\
		 & close to the Python style guide PEP-8~\cite{pep8:2001}. \\
Trouble  & pitfalls, mistakes, and how to avoid them.\\
         & also the place to look for (silent) assumptions.\\
\end{tabular}
\end{table}

All classes and all class methods have documentation strings that describe their use. A complete reference manual is extracted in https://bayesicfitting.nl, a site that still needs some work.

\begin{example*}[!ht]
\begin{tcolorbox}[bottomrule=0pt]
\par\noindent\rule{\textwidth}{0.8pt}
\refstepcounter{exctr}\label{ex:exo}
{\bf Example \arabic{exctr}: Boyle's Law}
\par\noindent\rule[6pt]{\textwidth}{0.4pt}
\begin{multicols}{2}
\setlength{\parindent}{10pt}

\label{sec:example}

The original data taken by Robert Boyle (1627-1691), that led to his eponymous law, have been handed down to us by \cite{magie:1935}. Boyle measured the pressure $p$ at various settings of a piston in a cylinder, resulting in the same amount of gas occupying different volumes, $V$. 

Boyle preceeded Bayes and Gauss and Laplace by a few generations. He did not know about least-squares and even less about model selection. His method was just the common sense that was later converted to numbers by Laplace.

Here we want to look at Boyle's data, considering 3 cases to see which case is more probable.

\subsection*{Case A}

The nominal relation between $V$ and $p$ is $pV = C$, or in terms of BayesicFitting we need a \Power\Model: $p = C / V$. The model has only 1 parameter $C$ for which we take a \Uniform\Prior\ from 0 to 50. The prior needs to be positive because pressure and volume are by nature positive quantities; so needs their product.

We define \NestedSampler, using a \Gauss\ErrorDistribution\ with unknown scale. This scale is a hyperparameter, that gets a \Jeffreys\Prior\ with limits between 0.01 and 1. 

\begin{Verbatim}[frame=single, fontsize=\small] 
 model = PowerModel( -1 )
 model.setLimits( 0, 50 )
 ns = NestedSampler( V, model, p )
 ns.distribution.setLimits( [0.01, 1] )
 evi = ns.sample( )
\end{Verbatim} 

Some key values of this solution are in table~E3.1 case A where the evidence, $\log_{10} Z$, the standard deviation of the residuals (scale), the log likelihood, and the parameters are listed. 
The fit itself is displayed in figure~E3.1. Although the fit seems reasonable, the residuals show a clear trend, with the largest residuals at the place where the model is steepest. 

\begin{center}
\includegraphics[width=0.4\textwidth]{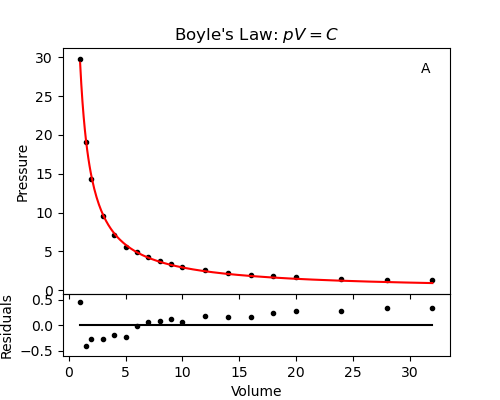}
\parbox{7cm}{\footnotesize Figure E3.1. Boyle's Law. The original data are in black. The best fit model is in red. In the lower panel the residuals are shown.}
\end{center}

Where the model is steep, a small error in $V$ entails a large error in $p$. As both values are readings from either a piston of a column of Hg, the errors could be similar, which inspired us to case B.

\subsection*{Case B}

As of the symmetry between $p$ and $V$ in Boyle's Law, $pV = C$, and the equivalent manner in which $p$ and $V$ are measured, it is hard to decide which one is the independent variable. We want to treat both variables similarly. We can use the \ErrorsInXandY\Problem. In this problem the values of the indepedent variable, $V$, need to be optimized too, and compared with the actually measured values, leading to N extra parameters to be estimated, one for every datapoint. These so-called nuisance parameters need a prior. We chose a Gaussian prior centered on the measured values with a width of 0.2, about the size of the noise scale in case A.

\begin{Verbatim}[frame=single, fontsize=\small]
 prb = ErrorsInXandYProblem( model, V, p )
 prb.prior = GaussPrior( scale=0.2 )
 ns = NestedSampler( problem=prb )
\end{Verbatim}

When we run the \NestedSampler, we get the values in table~E3.1 case B. The remaining noise has clearly gone down, especially in the steep regions. The likelihood has improved. However the evidence has not increased. Case B is less probable than case A by a factor 56 ($= 10^{5.09-3.34}$), see equation~\ref{eq:bayes-rev}. Even though case B has a higher likelihood, it is less probable than case A. This is due to the fact that it has much more wiggle room with all nuisance parameters required in this problem.

\begin{center}
\includegraphics[width=0.4\textwidth]{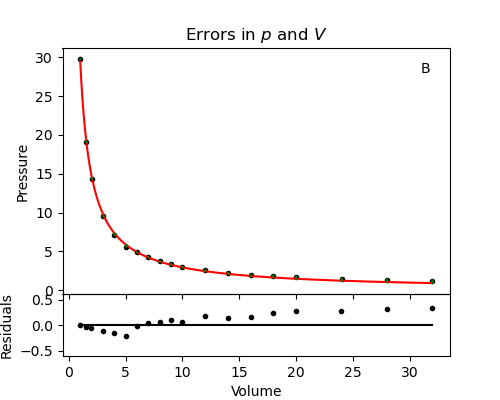}
\parbox{7cm}{\footnotesize Figure E3.2. Boyle's Law with similar sized errors in $p$ and $V$. }
\end{center}

\hfill\fbox{To be continued on the next page.}

\end{multicols}
\end{tcolorbox}
\end{example*}

\begin{example*}[!ht]
\begin{tcolorbox}[toprule=0pt]
\par\noindent\rule{\textwidth}{0.8pt}
{\bf Example \arabic{exctr}: Boyle's Law (continued)}
\par\noindent\rule[6pt]{\textwidth}{0.4pt}
\begin{multicols}{2}
\setlength{\parindent}{10pt}

\subsection*{Case C}

Closer inspection of the residuals in figure~E3.1, shows that the residuals are low on the left side and high on the right side. This suggests that we might have a zero point error in both variables. We define an adjusted law as
$(p-p_0)(V-V_0) = C$.

\begin{Verbatim}[frame=single, fontsize=\small]
 m1 = PolynomialModel( 0 )
 m1.setLimits( -2, 2 )
 m2 = PadeModel( 0, 1, fixed={2:1} )
 m2.setLimits( [0,-2], [50,2] ) 
 model = m1 + m2
\end{Verbatim}

\begin{center}
\includegraphics[width=0.4\textwidth]{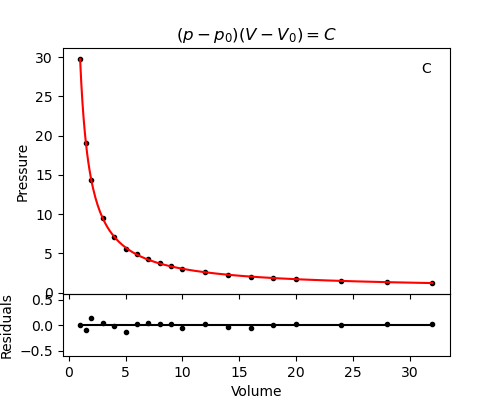}
\parbox{7cm}{\footnotesize Figure E3.3. Boyle's Law, recalibrated with zero point shift in $V$ and $p$.}
\end{center}

The $C$ parameter still has a uniform prior between 0 and 50. The two offsets have uniform priors between -2 and +2. 

Running this model with the same NestedSampler configuration, yield the values of table~E3.1, case C. The noise has decreased dramatically, resulting in a much higher \logL. This time, the evidence also increased significantly. Case C is more probable than case A by a factor of 13 million, again using equation~\ref{eq:bayes-rev}.

In figure~E3.3, the results for this adjusted model is shown. 
The fit is almost perfect; the residuals show no systematic effects. Not much more can be improved with this data.

\vspace{10pt}
\begin{center}
\begin{tabular}{|c|c|c|c|c|}
\hline
Case  &  $\log_{10} Z$  &  scale   &  logL & parameters\rule{0mm}{4mm} \\ 
\hline
A & -3.34 & 0.25 & -0.47 & $C = 29.30$\rule{0mm}{4mm} \\
B & -5.09 & 0.17 & 3.99 & $C = 30.05$ \\
C &  3.77 & 0.06 & 28.1 & $C = 26.20$ \\
  &       &      &      & $p_0 = 0.40$ \\
  &       &      &      & $V_0 = -0.11$ \\
\hline
\end{tabular}

\parbox{8cm}{\footnotesize \vspace{10pt} Table E3.1. Boyle's Law, recalibrated with zero point shift in $V$ and $p$.}

\end{center}

Although in hindsight and having only the data to judge by, it is impossible to say with certainty which of the cases reflects the truth. We can only state that case C is by far the most probable of the three.

\end{multicols}
\end{tcolorbox}
\end{example*}

\subsection{Exceptions}

Exceptions are only raised when the program has no way to succeed in its mission. E.g. when there are NaN's in its data, or when an attribute is changed in a way that makes the object inconsistent. Apart from that, the user has much leeway to (mis)use the toolbox. In the end, the user is responsible for the results.

The only exception defined in this toolbox is the \ConvergenceError\ that is raised when one of the \Fitter s does not converge properly. In all other cases the built-in errors are used

\subsection{Examples}

The examples in this paper are given to illustrate some points. They will not function as scripts to learn about the toolbox and to emulate upon. The BayesicFitting package contains a score of examples written in Jupyter notebook style. They are also in GitHub \cite{examples:2021} where they can be inspected and downloaded from.

\subsection{Tests}

A large package like this cannot be kept alive without test harnesses. We use the unittest testing environment, not in the last place because it was inspired by JUnit, the JAVA test environment that was used by HCSS. We could easily translate our test harnesses in JAVA to similar ones in PYTHON and expand from there. A good programmer is a lazy programmer, thinking before acting. 

All classes have their own tests. All tests have to be executed successfully, whenever a new upload is made to GitHub and PiPy. 

Strict regression tests, value by value the same as an earlier run, are hard for a package where random number generators (RNG) play a heavy role. A seed-initiated RNG will indeed produce the same sequence of random numbers. And a program using it, will produce the same results. However, when the program needs a bug correction or a feature extension that makes extra (or less) calls to the RNG too, the same sequence of random numbers is shifted with respect to the code and the program is send along a completely different path. The final results however should be the same within the statistical boundaries. That is what we test. It is less certain than strict regression but it is what it is.

\subsection{Math and Computation}

For all of the Bayesian Inference Theory used in this paper, the mathematics is mature and unproblematic. Unfortunately, the quantities that need to be handled, can be many orders of magnitude different in size, even when probabilities are conventionally done in logarithms. Our computers only know binary integers, sometimes disguised as floats, quasi-real numbers. 

All the computations are done in 64-bit floats, which have an accuracy of about 10 digits. Calculations with floats more than 10 orders different, quickly lose their meaning. As there are many opportunities to have numbers like that, in \Model s, likelihoods, \Prior s, \Fitter s, solving inference problems is as much parts science as it is art and craft and experience.

We have collected, mostly from our own experience, a number of pitfalls in a document, with possible ways to circumvent them in \cite{troubles:2021}.

\section{Summary}
\label{sec:summary}

We have described the BayesicFitting toolbox, with special emphasis on \NestedSampler\ and its pertaining classes. BayesicFitting is a PYTHON toolbox of over 100 classes that work together in a consistent way. Simple \Model\ classes can be combined in several ways to form compound models that are for all intents and purposes \Model s in their own right.

In a few lines, a simple inference problem can be stated and solved, in a Bayes-compatible fashion. Either using one of the many \Fitter s, getting a Gaussian approximation of the posterior and evidence, or using \NestedSampler\ which returns Bayes' evidence and posterior distribution of the parameters. More complicated problems take somewhat more lines but are still easily tractable. The toolbox has numerous examples to study and emulate.

\section*{Acknowledgments}

As many programmers know, it is easier to expand on existing code than to start from scratch. The BayesicFitting toolbox also shows some traces of external code. E.g. the \LevenbergMarquardt\Fitter\ and the \Amoeba\Fitter\ started as functions in Numerical Recipes \citep{press:1992}. And some code inside the \NestedSampler\ can be traced to an example program in \cite{sivia:2006}. All this code has seen at least 2 translations, from C to JAVA to PYTHON.

\bibliographystyle{elsarticle-num}
\bibliography{./hifi.bib} 

\end{document}